\documentclass[aps,prb,twocolumn,floatfix,amsmath,amssymb,superscriptaddress]{revtex4}
\usepackage[final]{graphicx}
\usepackage{bm}
\usepackage{afterpage}
\usepackage{color}
\usepackage{comment}
\usepackage[colorlinks=true,urlcolor=blue,linkcolor=blue,citecolor=blue]{hyperref}

\begin{document}

\title{Superconducting pairing symmetry in the kagome-lattice Hubbard model}

\author{Chenyue Wen$^*$}
\affiliation{School of Physics, Beihang University,
Beijing, 100191, China}

\author{Xingchuan Zhu\footnote{These authors contributed equally to this work} }
\email{xc\_zhu2017@mail.bnu.edu.cn}
\affiliation{Center for Basic Teaching and Experiment, Nanjing University of Science and Technology, Jiangyin 214443, People's Republic of China}

\author{Zhisong Xiao}
\affiliation{School of Physics, Beihang University,
Beijing, 100191, China}

\author{Ning Hao}
\affiliation{Anhui Key Laboratory of Condensed Matter Physics at Extreme Conditions, High Magnetic Field Laboratory, HFIPS, Anhui,
Chinese Academy of Sciences, Hefei, 230031, China}

\author{ Rubem Mondaini}
\affiliation{Beijing Computational Science Research Center, Beijing 100084, China}

\author{Huaiming Guo}
\email{hmguo@buaa.edu.cn}
\affiliation{School of Physics, Beihang University,
Beijing, 100191, China}
\affiliation{Beijing Computational Science Research Center, Beijing 100084, China}

\author{Shiping Feng}
\affiliation{ Department of Physics,  Beijing Normal University, Beijing, 100875, China}

\begin{abstract}
The dominating superconducting pairing symmetry of the kagome-lattice Hubbard model is investigated using the determinant quantum Monte Carlo method. The superconducting instability may occur when doping the correlated insulators formed by the Hubbard interaction near the Dirac filling, and the possible superconducting state exhibits an electron-hole asymmetry. Among the pairing symmetries
allowed, we demonstrate that the dominating channel is $d$-wave in the hole-doped case. This opens the possibility of condensation into an
unconventional $d_{x^2-y^2}+id_{xy}$ phase, which is characterized by an integer topological invariant and gapless edge
states. In contrast, the $s^*$-wave
channel, which has no sign change in the pairing function,
is favored by the electron doping. We further find the dominating $s^*$-wave pairing persists up to the Van Hove singularity. The results are closely related to the recent experimental observations in kagome compounds $\textrm{AV}_3\textrm{Sb}_5$(A: K, Rb,Cs), and provide insight into the pairing mechanism of their superconducting states.
\end{abstract}

\pacs{
  71.10.Fd, 
  03.65.Vf, 
  71.10.-w, 
}

\maketitle
\section{Introduction}
The kagome lattice, composed of corner-sharing triangles whose lattice points each have four nearest neighbors~\cite{mekata2003kagome}, combines the intriguing physics of geometry frustration, flat band, Van Hove singularity (VHS) and Dirac fermion, setting an ideal platform for novel quantum phases ~\cite{balents2010spin,Savary_2016,RevModPhys.89.025003,ye2018massive,kang2020dirac,PhysRevB.82.075125,PhysRevLett.103.046811,PhysRevB.90.035118,PhysRevB.85.144402,PhysRevB.87.115135,PhysRevB.94.014508,PhysRevLett.110.126405}.
Recently, a new family of kagome materials $\textrm{AV}_3\textrm{Sb}_5$(A: K, Rb, Cs) was discovered~\cite{ortiz2019}. They are composed of a layered structure, with an ideal kagome network of vanadium layers separated by alkali metal ions. The common properties of these compounds include: $Z_2$ topological metal state, charge density wave (CDW) order occurring below $T_c^{\rm CDW}\approx 80-110 \textrm{K}$, and unconventional superconductivity with critical temperature $T_c\approx 0.9-2.7\textrm{K}$~\cite{ortiz2020,jiang2021unconventional,Yin_2021,ortiz2021}.

Despite intense investigation on these compounds, debate over the nature of CDW and superconductivity persists~\cite{tan2021charge,FENG2021,zhou2021origin,wu2021nature,lin2021complex}. In particular, the controversy regarding the superconducting (SC) pairing and its mechanism abounds. While a significant residual linear term in the thermal conductivity, and the $V$-shaped spectral gap opening in the differential conductance indicate the unconventional nodal superconductivity~\cite{zhao2021nodal,chen2021roton}, the magnetic penetration depth and nuclear magnetic resonance measurements suggest a nodeless $s$-wave superconductor~\cite{duan2021nodeless,gupta2021microscopic,Mu_2021}. Furthermore, a recent scanning tunneling microscopy study in $\textrm{CsV}_3\textrm{Sb}_5$ finds evidence of the existence of gap nodes but the absence of sign change in the SC order parameter~\cite{xu2021multiband}. The disagreement among the different experimental setups is certainly influenced by the complexity induced by the multiband nature of the superconductivity in these compounds\cite{tsirlin2021anisotropic,kang2021twofold,luo2021electronic,hu2021rich,labollita2021tuning}. Since existing theoretical and experimental studies indicate that strong electron correlations play an essential role in the appearance of superconductivity~\cite{li2021observation,zhao2021electronic}, it is highly desirable to gain insight on the dominating pairing symmetry. For that, a good starting point is the Hubbard Hamiltonian on a kagome lattice.

In this paper, we employ the determinant quantum Monte Carlo method (DQMC)~\cite{PhysRevD.24.2278,Hirsch1985,PhysRevB.40.506,PhysRevB.39.839}, combined with cues coming from mean-field (MF) theory, to study the SC pairing symmetry in the kagome-lattice Hubbard model. The quasiparticle spectra of the pairing symmetries allowed on the kagome lattice are first analyzed to demonstrate properties of the different types of superconducting states. Then DQMC calculations reveal that doping around the Dirac point, the dominating pairing channel is $d$ ($s^*$)-wave in the hole (electron)-doped case, and the $s^*$-wave channel remains dominating even when the system is doped to the upper VHS. These results are consistent with some of recent experimental observations, and provide a theoretical understanding of the superconducting states in the newly discovered kagome materials.

\section{The model and method}

\begin{figure}[htbp]
\centering \includegraphics[width=6.5cm]{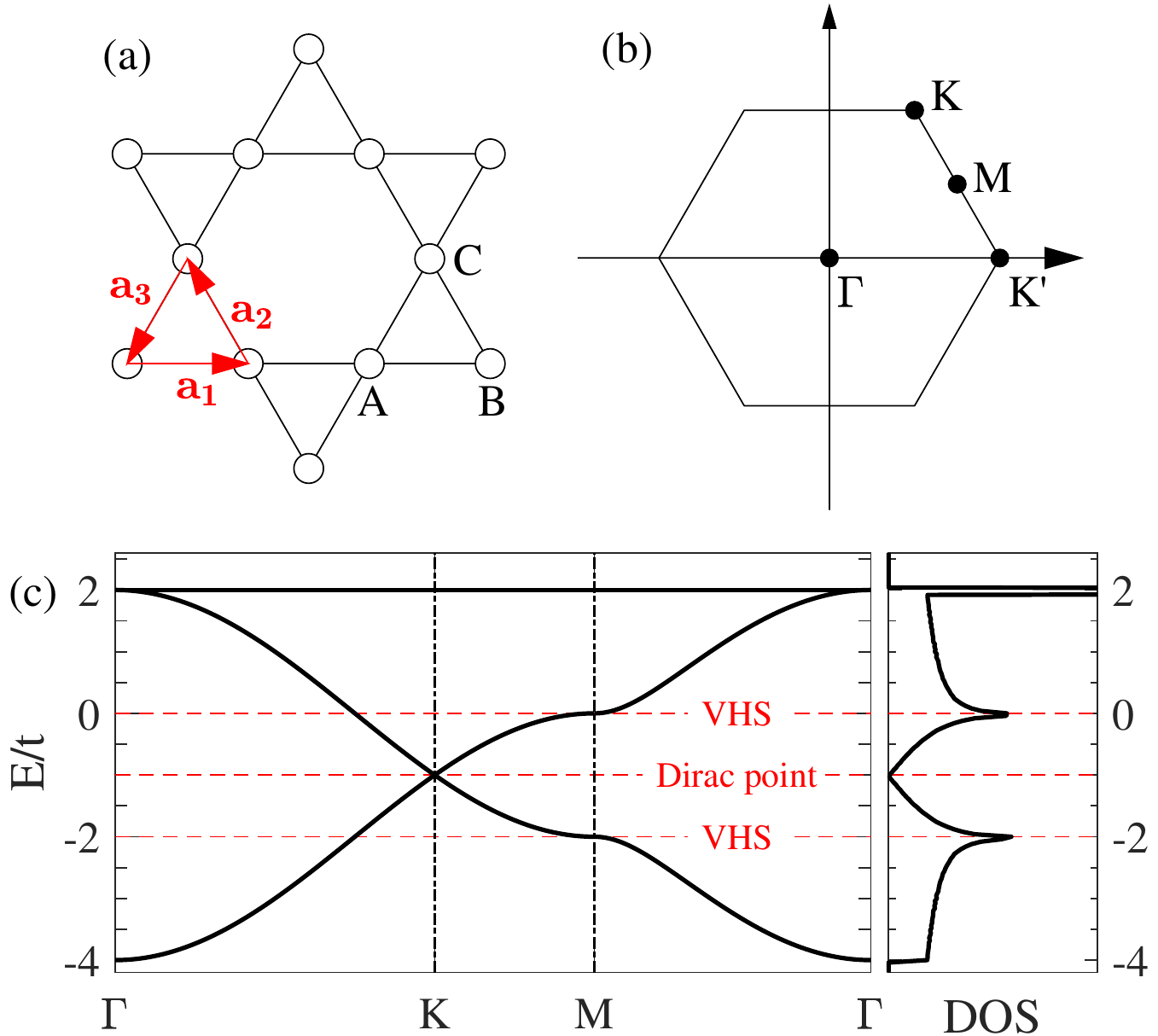} \caption{(a) The kagome lattice which is a triangular Bravais lattice with a three-site unit cell. The sublattice is labeled as A,B,C. (b) The first Brillouin zone with the high symmetry points marked. (c) The band structure along the high symmetry directions in the Brillouin zone (left) and the density of states (right).}
\label{fig1}
\end{figure}

We start from the kagome-lattice Hubbard model,
\begin{align}
H=-t\sum_{\langle i j\rangle \sigma} c_{i \sigma}^{\dagger} c_{j \sigma}^{\phantom{\dagger}}+U \sum_{i}\left(n_{i \uparrow}-\frac{1}{2}\right)\left(n_{i \downarrow}-\frac{1}{2}\right),
\label{eq:H}
\end{align}
where $c_{i \sigma}^{\dagger}$ and $c_{i \sigma}$ are the creation and annihilation operators, respectively, at site $i$ with spin $\sigma=\uparrow, \downarrow$; $\langle ij\rangle$ denotes nearest neighbors; $n_{i \sigma}=c_{i \sigma}^{\dagger} c_{i \sigma}$ is the number of electrons of spin $\sigma$ on site $i$, and $U$ is the on-site repulsion. Throughout the paper, the hopping amplitude is set to $t = 1$ as the unit of energy.

The kagome lattice has a three-site unit cell [Fig.~\ref{fig1}(a)]. In momentum space, the $U=0$ Hamiltonian is given by~\cite{guohm2009}
\begin{align}
\mathcal{H}_{0}({\mathbf{k}})=-2 t\left(\begin{array}{ccc}
0 & \cos k_{1} & \cos k_{3} \\
\cos k_{1} & 0 & \cos k_{2} \\
\cos k_{3} & \cos k_{2} & 0
\end{array}\right),
\end{align}
where $k_n={\mathbf{k}}\cdot {\mathbf{a}}_n$ (the sublattice index $n=1,2,3$) with ${\mathbf{a}}_1=(1,0),{\mathbf{a}}_2=(-1,\sqrt{3})/2$, and ${\mathbf{a}}_3=-({\mathbf{a}}_1+{\mathbf{a}}_2)$. The spectrum of $\mathcal{H}_{0}({\mathbf{k}})$ has one flat band $E_{3}({\mathbf{k}})=2t$ and two dispersive ones,
\begin{align}
E_{1,2}({\mathbf{k}})=t[-1\pm \sqrt{4f({\mathbf{k}})-3}],
\end{align}
with $f({\mathbf{k}})=\cos^2 k_1+\cos^2 k_2+\cos^2 k_3$. Bands $1$ and $2$ touch at two inequivalent Dirac points ${\bf K}_{\pm}=(\pm 2\pi/3,0)$ at energy $-t$, see Figs.~\ref{fig1}(b) and \ref{fig1}(c). For $\frac{1}{3}$ filling, the lowest band is filled, and the low-energy excitations resemble those of graphene, which are linear, $\epsilon_{1,2}=\pm \sqrt{3}t|\vec{q}|$, with $\vec{q}=(q_x,q_y)$ a small displacement away from the Dirac points.

For the dispersive bands, three momenta $M$ at the centers of the edges of the Brillouin zone (BZ) are saddle points, resulting in VHSs at fillings $\rho=1/4$ and $5/12$, respectively. The corresponding energies are $E_{M}/t=-2$ and $0$, and the Dirac points are exactly located at the middle point between them.

Conversely, at finite interactions, Eq.\eqref{eq:H} is solved numerically via DQMC, where one decouples the on-site interaction term through the introduction of an auxiliary Hubbard-Stratonovich field, which is integrated out stochastically. The only errors are those associated with the statistical sampling, the finite spatial lattice size, and the inverse temperature discretization. These errors are well controlled in the sense that they can be systematically reduced as needed, and further eliminated by appropriate extrapolations. Since the kagome lattice is non-bipartite,  the infamous sign problem exists at all densities, and can become severe upon lowering the temperature and increasing the interaction strength~\cite{PhysRevB.41.9301,PhysRevLett.94.170201,PhysRevB.92.045110}. Nevertheless, the sign problem is found to be significantly reduced at some specific fillings, where the simulations can be carried out at relatively low temperatures. Generally, we access the temperatures with the average sign higher than $0.5$ to obtain reliable results. In the following, we use the inverse temperature discretization $\Delta\tau=0.1$, and lowest temperature accessed is $T/t=1/25$. The lattice has $N=3\times L \times L$ sites with $L$ up to $9$.

\section{Pairing symmetries and properties of the SC state}
\begin{figure}[htbp]
\centering \includegraphics[width=7.5cm]{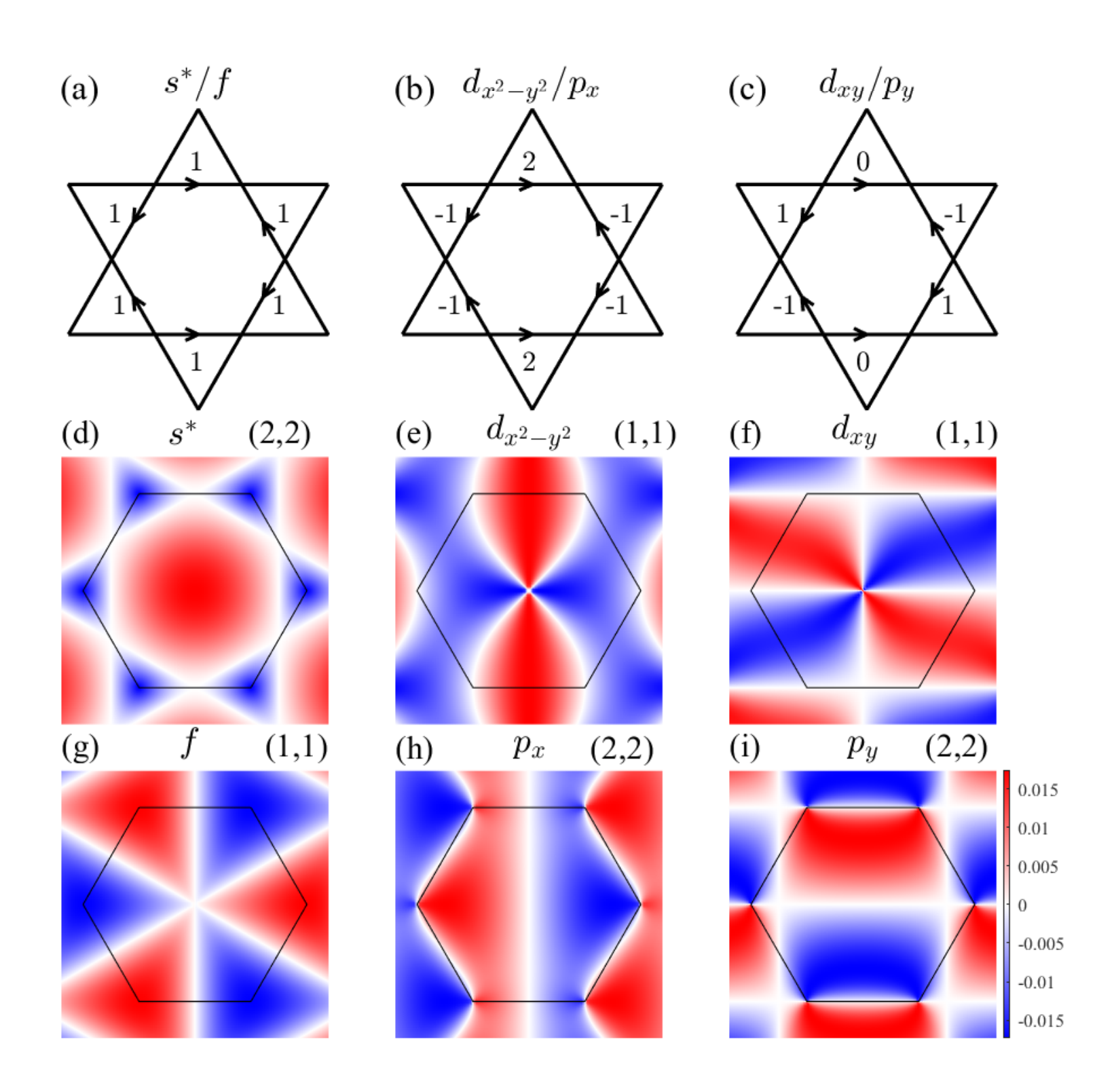} \caption{The pairing symmetries allowed by the point group of the kagome lattice: (a) $s^*$- and $f$-wave, (b) $d_{x^2-y^2}$- and $p_x$-wave, and (d) $d_{xy}$- and $p_y$-wave. For triplet pairing, there is an additional sign when the pairing is along the opposite direction of the arrow. The momentum dependence of the intra-band pairing function for (d) $s^*$, (e) $d_{x^2-y^2}$, (f) $d_{xy}$, (g) $f$, (h) $p_x$, and (i) $p_y$ pairing channels.}
\label{fig2}
\end{figure}

The on-site repulsive interactions tend to drive nonlocal pairings, and hence nearest-neighbor SC pairings are considered. Their symmetries should be compatible with the underlying geometry of the lattice. As the kagome lattice is described by $C_{6v}$ point group symmetry, the possible pairing states can be classified by the irreducible representations of $C_{6 v}$. These include the singlet pairing symmetries: $s^{*}$-wave, $d_{x^{2}-y^{2}}$-wave, and $d_{x y}$-wave; and triplet pairing symmetries: $p_{x}$-wave, $p_{y}$-wave, and $f$-wave, all of which are schematically represented in Fig.~\ref{fig2}. While $s^{*}$- and $f$-wave correspond to one-dimensional representations,
$d_{x^2-y^2}$ and $d_{xy}$ belong to two-dimensional representation $E_2$, and thus are degenerate, which also holds for any linear combination of them. The value of the pairing susceptibility is the same for all the degenerate combinations, thus can not distinguish them from one another. A qualitative argument for the dominance of a chiral combination, i.e., $d+id$ pairing, is that it opens a gap everywhere in the spectrum. Compared to the individual $d_{x^2-y^2}$ or $d_{xy}$ pairing which has nodes in the spectrum, the gap opening enable an overall energy lowering, hence makes the $d+id$ pairing energetically favored. Similarly, $p_{x}$ and $p_{y}$ belong to two-dimensional representation $E_{1}$, and the above argument is also applicable to the formation of chiral $p+ip$ superconductivity if the $p$-wave channel dominates.

We then explore the properties of the SC state with the above allowed pairing symmetries based on the
BCS Hamiltonian, which writes as
\begin{align}
H_{\rm SC}=\sum_{\mathbf{k}} \Psi_{\mathbf{k}}^{\dagger} \mathcal{H}_{\mathbf{k}} \Psi_{\mathbf{k}},
\label{eq:H_SC}
\end{align}
with $\Psi_{\mathbf{k}}=\left(c_{1, \mathbf{k} \uparrow}, c_{2, \mathbf{k} \uparrow}, c_{3, \mathbf{k} \uparrow}, c_{1,-\mathbf{k} \downarrow}^{\dagger}, c_{2,-\mathbf{k} \downarrow}^{\dagger}, c_{3,-\mathbf{k} \downarrow}^{\dagger}\right)^{T}$ and
\begin{align}
\begin{aligned}
\mathcal{H}_{\mathbf{k}} &=\left[\begin{array}{cc}
\mathcal{H}_{0}(\mathbf{k})-\mu & \Delta_{\mathbf{k}}^{\dagger} \\
\Delta_{\mathbf{k}} & -\mathcal{H}_{0}(\mathbf{k})+\mu
\end{array}\right], \\
\Delta_{\mathbf{k}} &=\left[\begin{array}{ccc}
0 & \eta_{1}({\mathbf{k}}) & \eta_{3}({\mathbf{k}}) \\
\eta_{1}({\mathbf{k}}) & 0 & \eta_{2}({\mathbf{k}}) \\
\eta_{3}({\mathbf{k}}) & \eta_{2}({\mathbf{k}}) & 0
\end{array}\right] .
\end{aligned}
\end{align}
Here $\mu$ is the chemical potential. $\eta_{n}({\mathbf{k}})=-\Delta_{n}(e^{-ik_n}-\zeta e^{ik_n})$ with the pairing amplitude $\Delta_{n}$, which
can be read from the real space arrangement in Fig.~\ref{fig2}; $\zeta=-1(+1)$ for singlet (triplet) pairing.

To demonstrate the symmetry of each pairing channel, the SC Hamiltonian $\mathcal{H}_{\mathbf{k}}$ should be transformed into the band basis, under which $\mathcal{H}_{0}(\mathbf{k})$ becomes diagonal, i.e., $V_{\mathbf{k}}^{\dagger}\mathcal{H}_{0}(\mathbf{k})V_{\mathbf{k}}=\textrm{diag}[E_{1}({\mathbf{k}}),E_{2}({\mathbf{k}}),E_{3}({\mathbf{k}})]$, where $V_{\mathbf{k}}$ is the eigenvector matrix of $\mathcal{H}_{0}(\mathbf{k})$. In turn, the pairing matrix becomes $\tilde{\Delta}_{\mathbf{k}}=V_{\mathbf{k}}^{\dagger} \Delta_{\mathbf{k}} V_{\mathbf{k}}$, and is composed of inter- and intra-band pairings.
As shown in Fig.~\ref{fig2}, the symmetries of the different pairing channels are clearly reflected in the intra-band pairings.

\begin{figure}[htbp]
\centering \includegraphics[width=7.5cm]{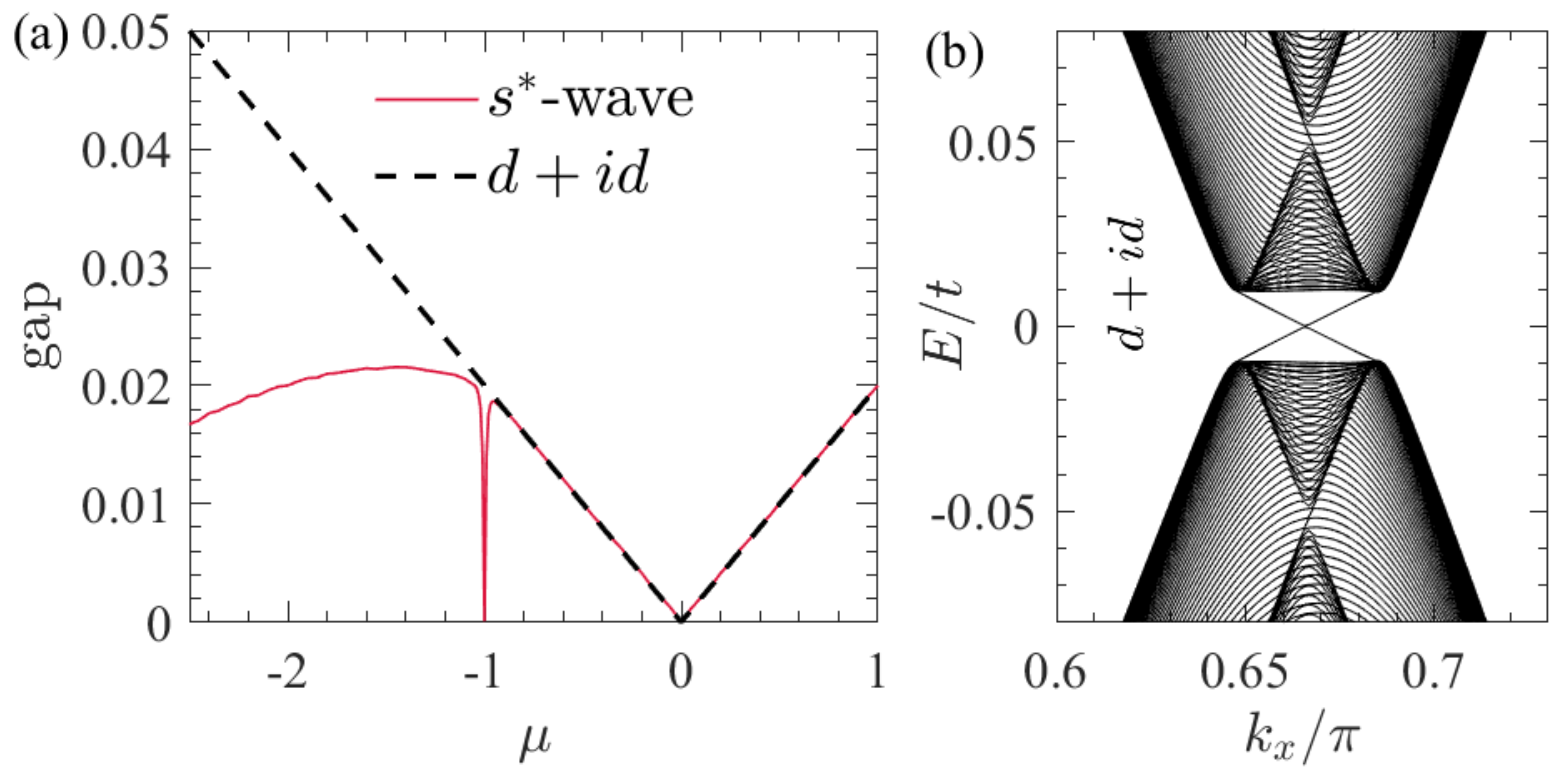} \caption{(a), The quasiparticle energy gap as a function of the chemical potential for the $s^*$-wave and $d + {\rm i}d$ superconductors. (b), The quasiparticle spectrum of the $d + {\rm i}d$ chiral superconducting state on kagome nanoribbons with a pair of sawtooth edges. The spectrum is symmetric to $k_x=\pi$, and we only show the low-energy part near the left valley. The gap parameter is $\Delta=0.01$. In (b), $\mu=-0.95$ corresponds to an electron doping on the Dirac cones. }
\label{fig3}
\end{figure}

In the presence of the complex inter-band pairings, the quasiparticle spectrum does not follow the standard BCS form, and it is not straightforward to identify whether there are zero-energy quasiparticles. Instead, the Hamiltonian in Eq.~\eqref{eq:H_SC} should be investigated via numerical diagonalization. We first focus on the case of the electron doping on the Dirac cones. It is found that the $s^{*}$ and $f$-wave states are fully gapped. Although the $d_{x^2-y^2}$- and $d_{xy}$-wave pairings are gapless, a
chiral linear combination of them, i.e., $d_{x^2-y^2}+{\rm i}d_{xy}$, is gapped. The
chiral $d$-wave state is a topological superconductor characterized by an integer Chern number $C=-2$, corresponding to which two pairs of gapless states traversing the gap appear in the presence of sawtooth edges [see Fig.~\ref{fig3}(b)]. The $p$-wave pairings are similar, and the chiral $p_x+{\rm i}p_y$-wave
state is also a topological superconductor with $C = -2$.

Figure \ref{fig3}(a) plots the quasiparticle energy gap as a function
of the chemical potential for the $s^*$-wave and $d + {\rm i}d$
superconductors. The gap of the $d+{\rm i}d$ state closes at the Dirac points ($\mu=-1$), where the Chern number changes its sign. For both channels, the gap is asymmetric to the Dirac points, and continuously decreases as $\mu$ increases. Above the Dirac points, both pairing channels have almost identical gap values, and become nodal superconductors at the upper VHS. In contrast, the gap of the triplet $p+{\rm i}p$ state closes at both the Dirac points and the two VHSs, and the Chern number always changes the value and sign across the gapless points (see Appendix B).

\section{DQMC study of the dominating pairing symmetry}

The low energy physics at density $\rho=2/3$ should be compared to the one of the honeycomb Hubbard model at half-filling ($\rho=1$), which undergoes a Dirac semi-metal to insulator transition at sufficiently large interactions. As we are interested in the SC instabilities, the best candidates are when doping away from this regime. We investigate densities symmetrically around the Dirac point, $\rho = 2/3 \pm \delta\rho$ (with $\delta\rho \simeq 0.046$).
To determine the dominating pairing symmetry, we evaluate the uniform pairing susceptibility~\cite{PhysRevB.91.241107,PhysRevB.97.155146,PhysRevB.97.235453},
\begin{equation}
   \chi^{\alpha}=\frac{1}{N} \int_{0}^{\beta} d \tau \sum_{i j}\left\langle\Delta_{i}^{\alpha}(\tau) \Delta_{j}^{\alpha \dagger}(0)\right\rangle,
\end{equation}
where $ \Delta_{i}^{\alpha}(\tau)= \sum_{j} f_{i j}^{\alpha} e^{\tau H} c_{i \uparrow} c_{j \downarrow} e^{-\tau H}$ is the time-dependent pairing operator with form-factors $f_{i j}^{\alpha}=0, \pm 1 \text { or } \pm 2$ for the bond
connecting sites $i$ and $j$, depending on the pairing symmetry $\alpha$ (see Fig.~\ref{fig2}). The effective susceptibility, $\chi_{\text {eff }}^{\alpha}\equiv\chi^{\alpha}-\chi_{0}^{\alpha}$, subtracts the uncorrelated part $\chi_{0}^{\alpha}$ from $\chi^{\alpha}$, thereby directly capturing the interaction effects, and can be further used to evaluate the pairing vertex.

\begin{figure}[htbp]
\centering \includegraphics[width=7.cm]{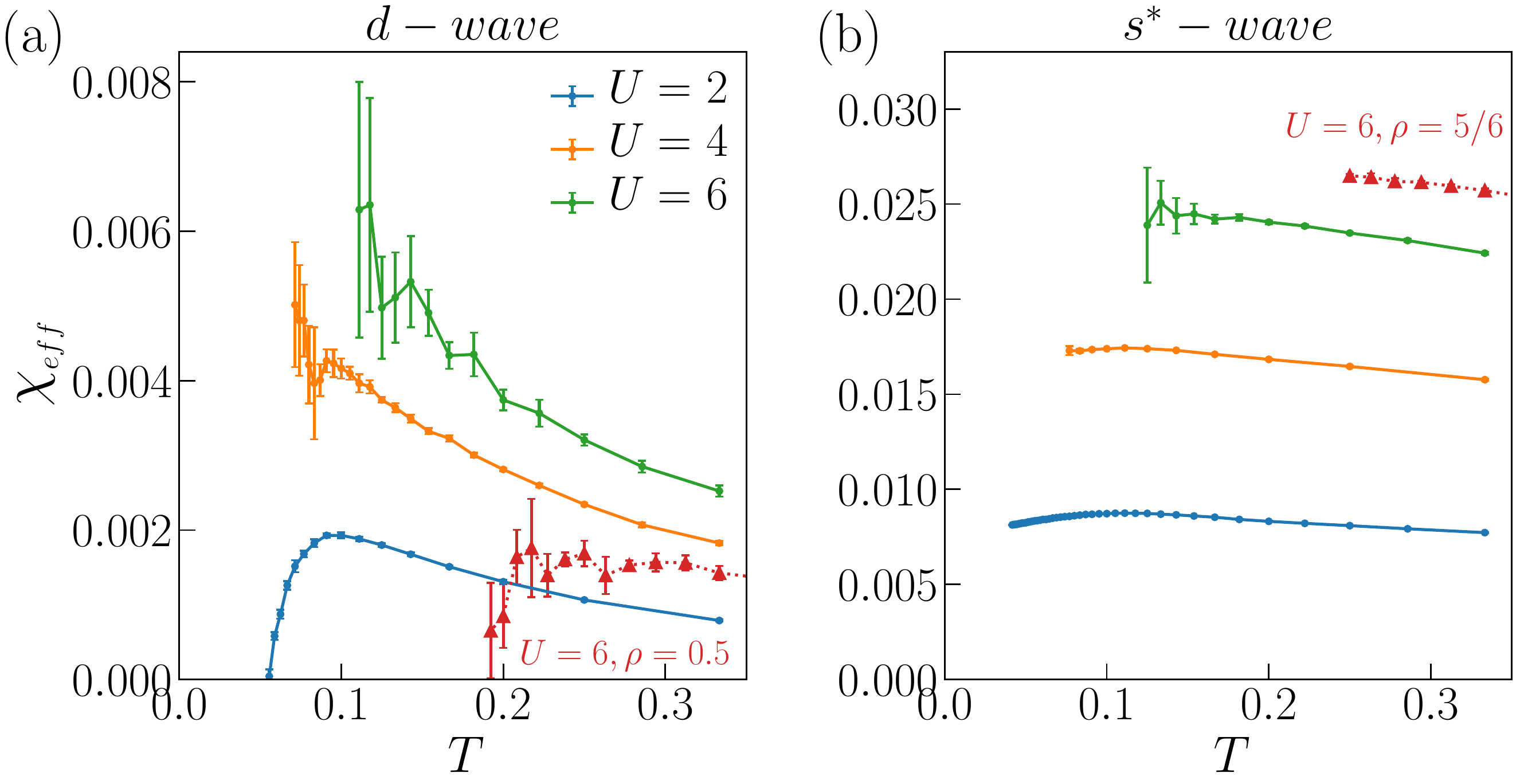} \caption{The effective susceptibility of the dominating pairing channel as a function of temperature for several values of $U$: (a) $d$-wave at $\rho_{_1}=0.62$; (b) $s^*$-wave at $\rho_{_2}=0.713$. Here $\rho_{_1}, \rho_{_2}$ are symmetric to the Dirac points, and correspond to hole and electron dopings, respectively. $\chi^{d}_{\rm eff}$ ($\chi^{s^*}_{\rm eff}$) at the lower (upper) VHS is also plotted in the left (right) panel to demonstrate the evolution with doping.}
\label{fig4}
\end{figure}

Figure \ref{fig4} shows $\chi^{}_{\rm eff}$ versus temperature for the pairing channels, $d$-wave at $\rho_{_1}=0.62$ and $s^*$-wave at $\rho_{_2}=0.713$, for several values of $U$. $\chi_{\rm eff}^{d}$ in (a) and $\chi_{\rm eff}^{s^*}$ in (b) are positive, and tend to increase rapidly for large $U$ at low temperatures.
In contrast, the values for triplet $p$-wave and $f$-wave pairings are increasingly negative with decreasing the temperature and increasing the interaction, suggesting these symmetry channels are suppressed (see Appendix D). These results demonstrate that the possible SC states on the upper and lower sides of the Dirac points are asymmetric: While the dominating pairing symmetry is $d$-wave in the hole-doped case, the $s^*$-wave channel, which has no sign change in the pairing function, is favored by the electron doping on the Dirac points. As the fillings approach the VHSs, the sign problem gets continuously worse, and the simulations are limited to relatively high temperatures. Although the values of $\chi^{d,s^*}_{\rm eff}$ still dominate at both VHSs, Fig.~\ref{fig4}(a) shows that $\chi^{d}_{\rm eff}$ drops quickly, and becomes negative at low temperatures, implying the possible SC instability is destructed in the heavily hole-doped case. In comparison, the $s^*$-wave pairing is successively enhanced by the electron doping, and remains dominating even at the upper VHS [Fig.~\ref{fig4}(b)]. Here the temperatures accessed by DQMC is still much higher than the SC transition temperature. Since the long-range SC order is still under development, the pairing susceptibility does not show significant finite-size effect.

\begin{figure}[htbp]
\centering \includegraphics[width=7.cm]{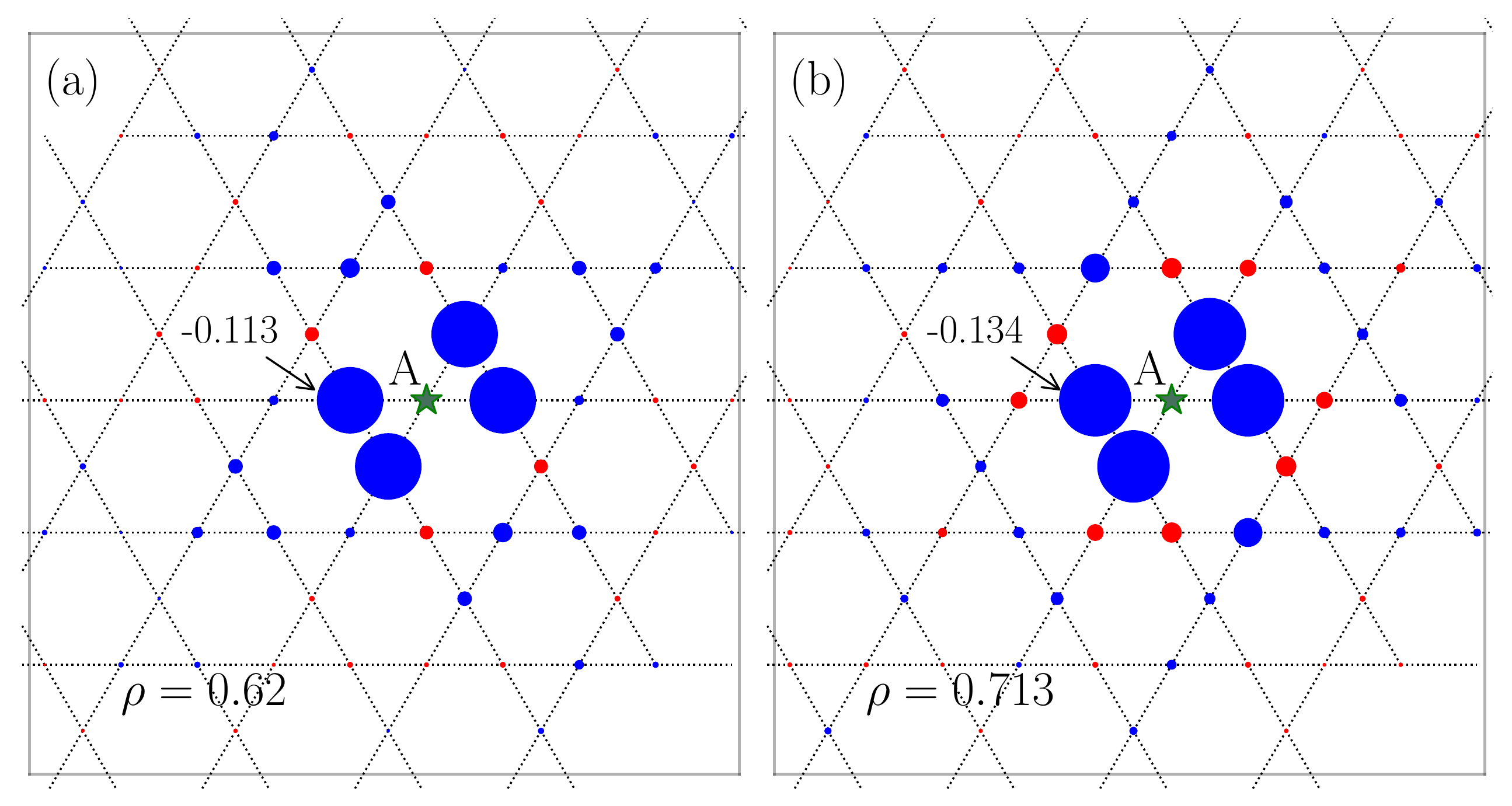} \caption{The spin-spin correlation at: (a) $\rho=0.62$; (b) $\rho=0.713$. The star symbol marks the reference site, and here a site in Sublattice A is chosen. The magnitude of the correlation is represented by the radii of the solid circle. The blue (red) color corresponds to the negative (positive) sign. The interaction strength is $U/t=4$, and the inverse temperature is $\beta t=6$.}
\label{fig5}
\end{figure}

To reveal the microscopic origin of the SC pairing interaction, the charge and spin correlations are calculated at both densities $\rho=0.62$ and $\rho=0.713$. No clear difference in the density-density correlations between the two fillings is observed under the same conditions (see Appendix D). The spin-spin correlations, on the other hand, are quite distinct. Specifically, at both fillings, the nearest-neighbor and next-nearest-neighbor spin correlations are antiferromagnetic and ferromagnetic, respectively. However the next-next-nearest-neighbor ones are antiferromagnetic (ferromagnetic) for $\rho=0.62$ ($\rho=0.713$). This is consistent with the asymmetry of the SC states below and above the Dirac points, which suggests superconductivity may be stimulated (and coupled) by magnetic fluctuations.

\section{Conclusions}
We have studied the SC pairing symmetry of the kagome-lattice Hubbard model using the DQMC method. From the high temperature trends of the pairing susceptibility, we find the possible superconducting states are asymmetric on the hole- and electron-doped sides of the Dirac points. While the dominating channel is $d$-wave for the hole doping, the $s^*$-wave pairing is favored in the electron doped regime. Besides, the $s^*$-wave symmetry channel remains dominating even when the system is doped to the upper VHS. We find the spin-spin correlation exhibit a similar electron-hole asymmetry, suggestive that the SC pairing interaction may be connected to magnetic fluctuations.

Experimentally, angle-resolved photoemission spectrum and other techniques have revealed that the Fermi surface contains pockets formed by different kinds of bands, and the superconductivity is of multi-band nature. Although a description of the electronic structure using a simple one-band model is often insufficient, the predicted $s^*$-wave pairing symmetry near the VHS is consistent with the experimental results~\cite{duan2021nodeless,gupta2021microscopic,Mu_2021,xu2021multiband}. Thus our results imply that the kagome-lattice Hubbard Hamiltonian includes some of the key ingredients that may explain the superconductivity in the family of kagome materials $\textrm{AV}_3\textrm{Sb}_5$(A: K, Rb,Cs), and further suggest the role of spin excitations in influencing the pairing interactions.

Although the on-site Hubbard repulsion can produce results consistent with some experimental observations, long-range interactions may not be omitted in real kagome superconductors. A recent theoretical study finds that the non-local Coulomb repulsion is a critical parameter to determine the pairing symmetry\cite{wu2021nature}. Besides, an exotic SC state, known as pair density wave, is revealed in recent STM experiments\cite{chen2021roton}. Hence simulating long-range interactions with the aim to explain the intriguing phenomena in kagome materials will be a significant direction of future DQMC research.

\section*{Acknowledgments}
The authors thank Fan Yang and Wen Yang for helpful discussions. J.S and H.G. acknowledge support from the National Natural Science Foundation of China (NSFC) grant Nos.~11774019 and 12074022, the NSAF grant in NSFC with grant No. U1930402, the Fundamental Research
Funds for the Central Universities and the HPC resources
at Beihang University. Z. X. acknowledges support from NSFC grant No.~61975005. R.M. acknowledges support from NSFC grants No.~U1930402, 12050410263, 12111530010 and No.~11974039.
N.H. acknowledges support from NSFC Grants
No. 12022413, No. 11674331, the “Strategic Priority Research Program (B)” of the Chinese
Academy of Sciences, Grant No. XDB33030100, the ‘100 Talents Project’of the Chinese Academy of Sciences, the Collaborative Innovation Program of Hefei Science Center, CAS (Grants No. 2020HSC-CIP002), the CASHIPS Director's Fund (BJPY2019B03).
S.F.~is supported by the National Key Research and Development Program of China, and NSFC under Grant Nos.~11974051 and 11734002.


\appendix


\renewcommand{\thefigure}{A\arabic{figure}}

\setcounter{figure}{0}

\section{The momentum dependence of the pairing functions}

\begin{figure}[htbp]
\centering \includegraphics[width=7.5cm]{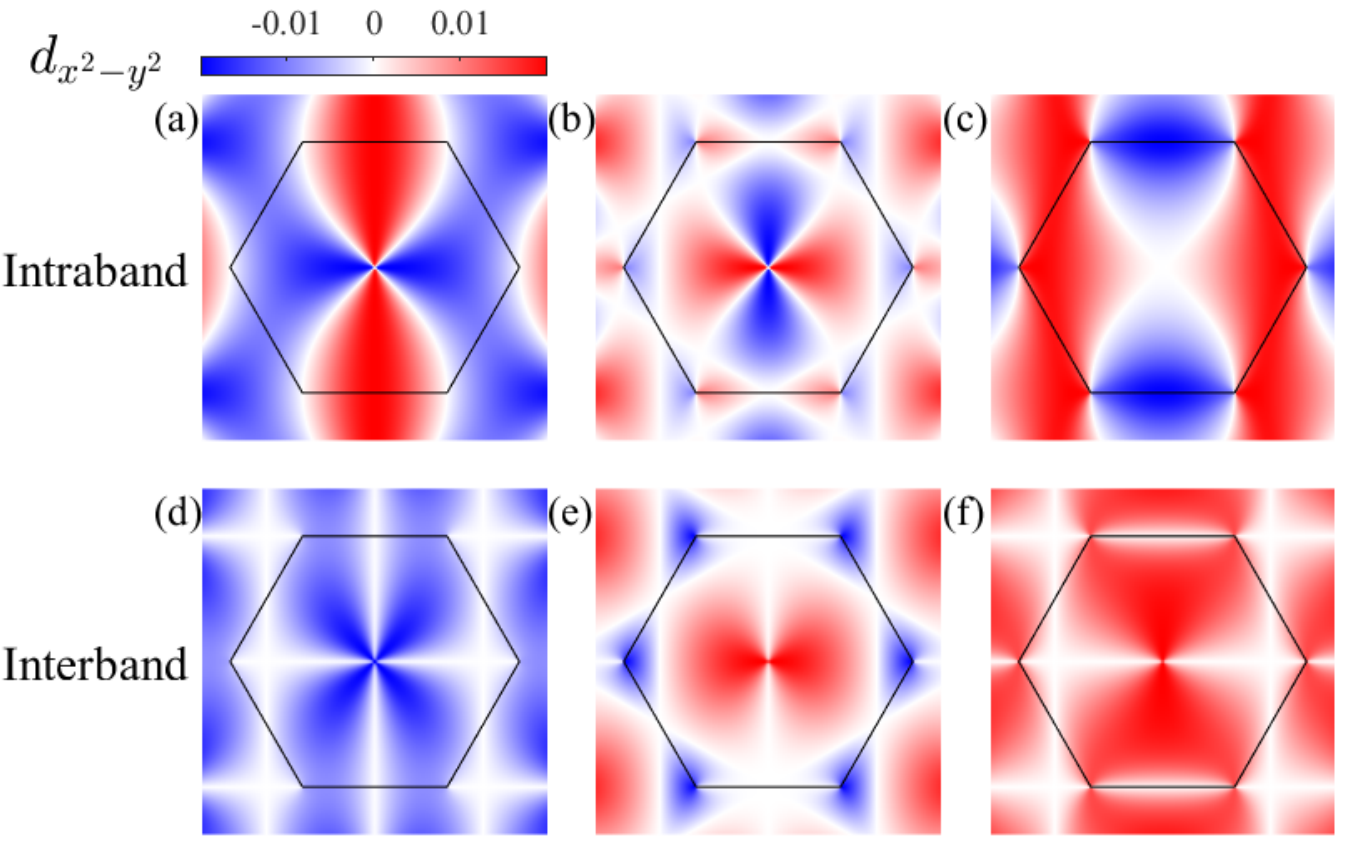} \caption{The momentum dependence of the $d_{x^2-y^2}$-wave intraband pairing function for the component (a) $(1,1)$, (b) $(2,2)$, (c) $(3,3)$. The interband pairing function for the component (d) $(1,2)$, (e) $(1,3)$, (f) $(2,3)$. Here $i,j=1,2,3$ in $(i,j)$ represent the band index.}
\label{figA1}
\end{figure}

\begin{figure}[htbp]
\centering \includegraphics[width=7.5cm]{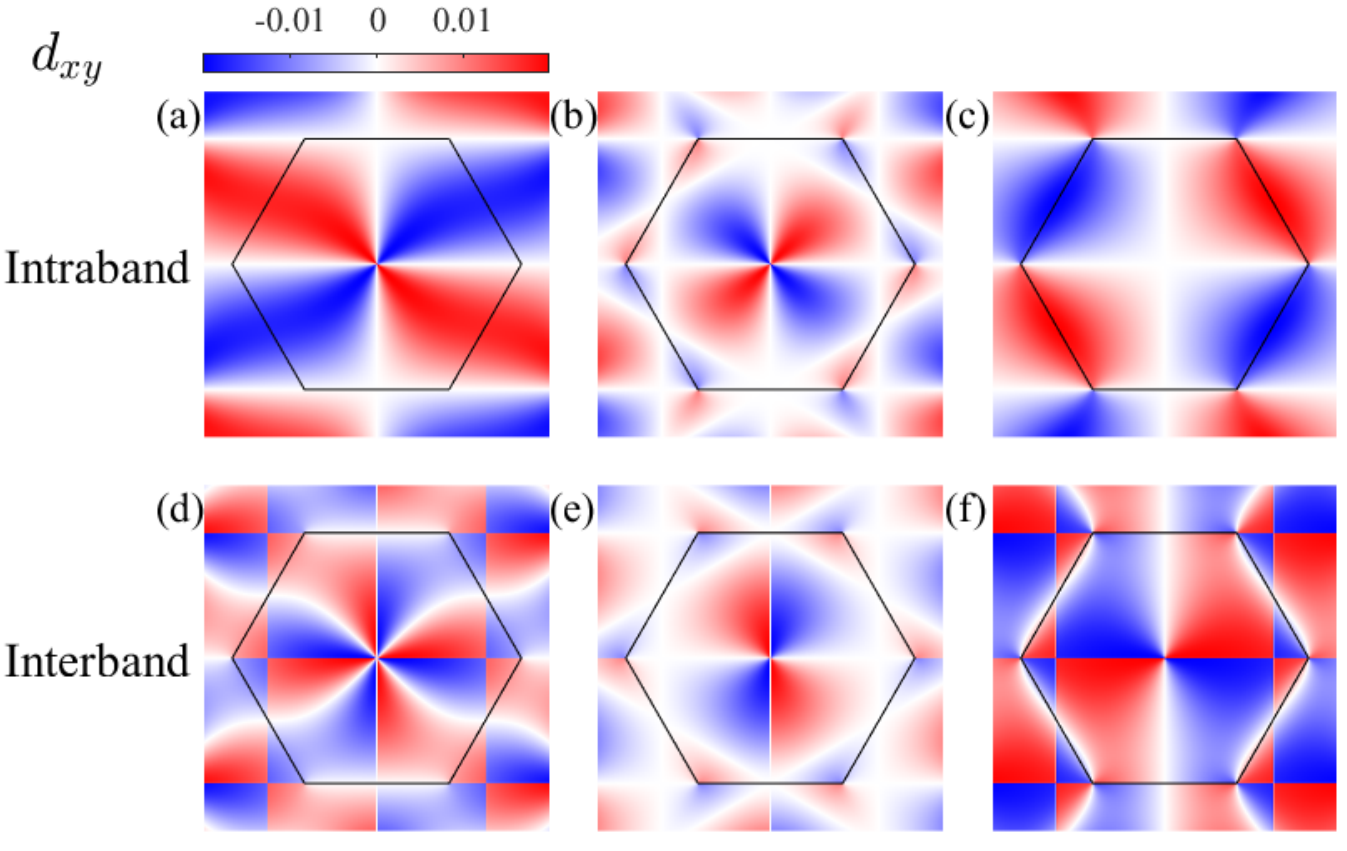} \caption{The momentum dependence of the $d_{xy}$-wave intraband pairing function for the component (a) $(1,1)$, (b) $(2,2)$, (c) $(3,3)$. The interband pairing function for the component (d) $(1,2)$, (e) $(1,3)$, (f) $(2,3)$.}
\label{figA2}
\end{figure}

\begin{figure}[htbp]
\centering \includegraphics[width=6.5cm]{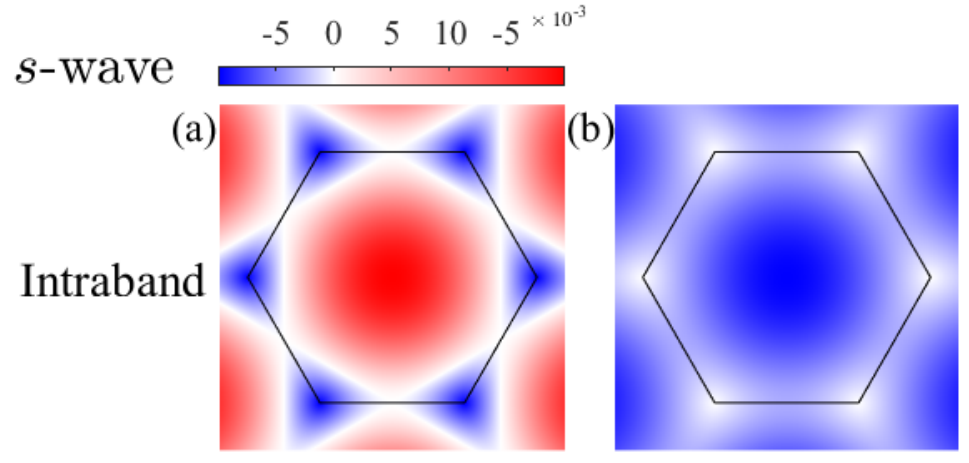} \caption{Momentum dependence of the $s$-wave intraband pairing function for the component (a) $(2,2)$, (b) $(3,3)$. All the other elements vanish for the $s$-wave channel.}
\label{figA3}
\end{figure}

The pairing function in the band basis is a $3\times 3$ matrix, in which the diagonal (off-diagonal) elements correspond to intra-band (inter-band) pairings. Figure \ref{figA1} plots the momentum dependence of the $d_{x^2-y^2}$-wave pairing function. Since the pairing matrix is symmetric, only nonequivalent elements are presented. As shown in Fig.~\ref{figA1}, the symmetry of the pairing function is clearly reflected in the intra-band pairings. The pairing functions for the $d_{xy}$-wave and $s$-wave are shown in Fig.~\ref{figA2} and Fig.~\ref{figA3}, respectively. The triplet pairing functions can be similarly obtained, and are not plotted here.

\section{More results on the quasiparticle spectrum}

\begin{figure}[htbp]
\centering \includegraphics[width=6.5cm]{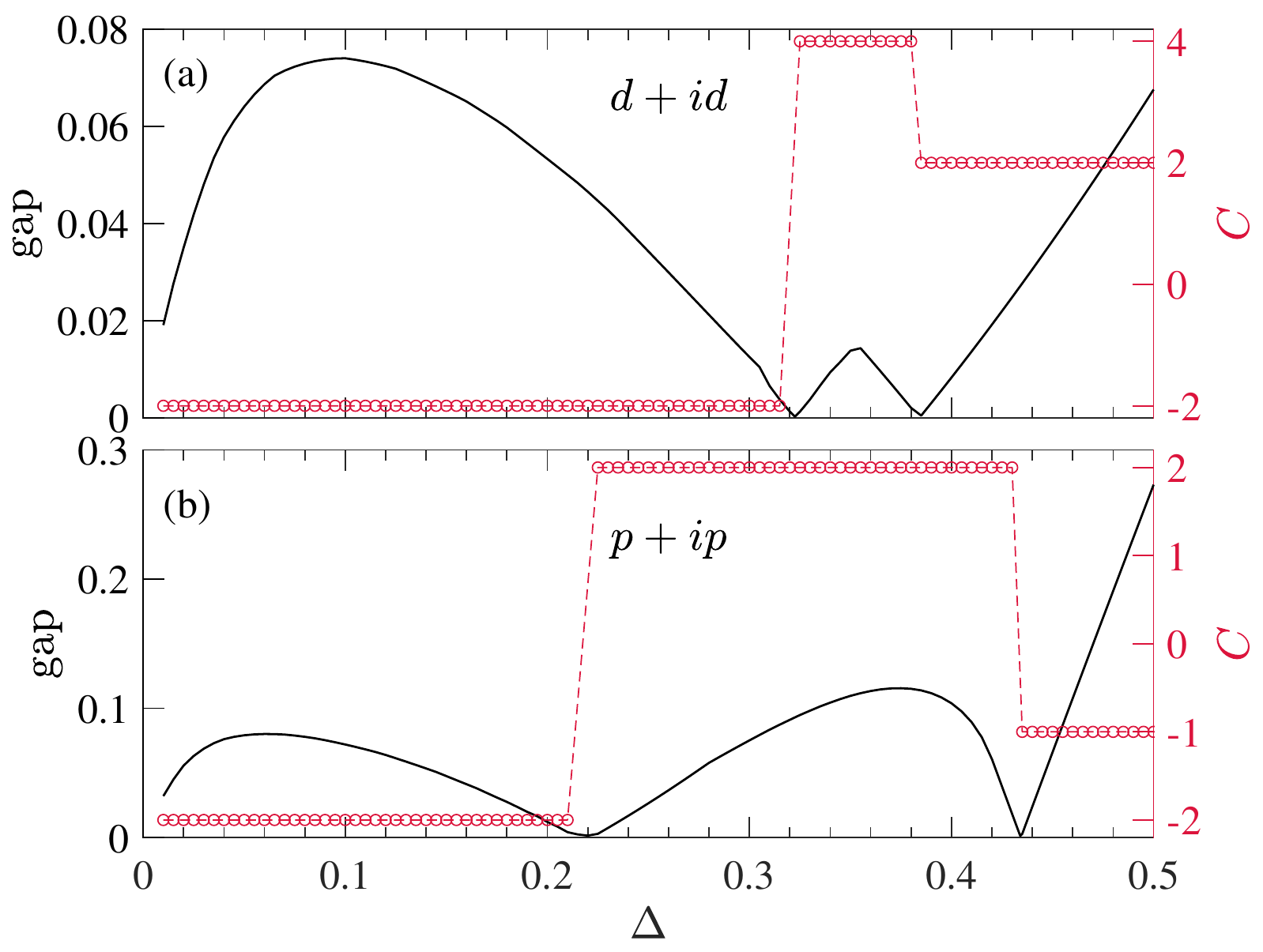} \caption{(a), The quasiparticle energy gap and the Chern number as a function of the pairing strength $\Delta$ for: (a) $d+{\rm i}d$ and (b) $p + {\rm i}p$ superconductors. Here $\mu=-0.95$ corresponds to an electron doping on the Dirac cones.}
\label{figa4}
\end{figure}

The quasiparticle spectrum evolves with the pairing strength $\Delta$. Figure \ref{figa4} plots the quasiparticle energy gap as a function of $\Delta$ for $d+{\rm i}d$ and $p+{\rm i}p$ superconductors. For both pairing channels, the gap closes two times in the calculated range $0<\Delta<0.5$, accompanied by a change of the Chern number. Between two gapless points, the gap value exhibits a dome-like shape. In practice, $\Delta$ should be a small fraction of the energy scale $t$. Thus the gap should be within the first dome, and both the $d+{\rm i}d$ and $p+{\rm i}p$ states are nodeless superconductors.

\begin{figure}[htbp]
\centering \includegraphics[width=6.5cm]{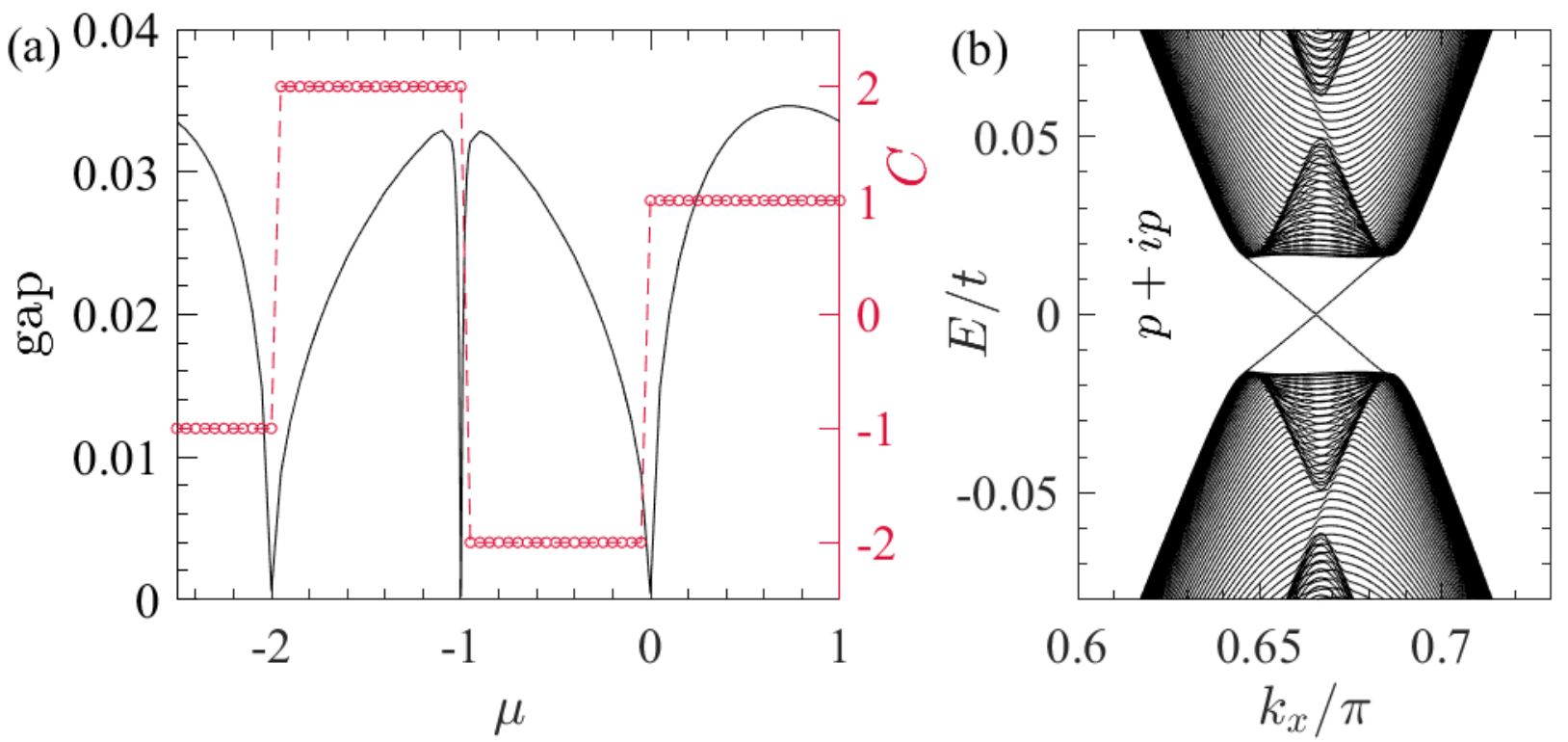} \caption{(a), The quasiparticle energy gap and the Chern number as a function of the chemical potential for the $p + {\rm i}p$ superconductor. (b), The quasiparticle spectrum of the $p + {\rm i}p$ chiral superconducting state on kagome nanoribbons with a pair of sawtooth edges. This plot is similar to Fig.~3 of the main text and the same parameters are used.}
\label{figA5}
\end{figure}

Figure \ref{figA5}(a) plots the quasiparticle energy gap and the Chern number as a function of the chemical potential for the $p + {\rm i}p$ state.  The gap of the triplet $p + {\rm i}p$ state closes at both the Dirac points and VHSs, and the Chern number always changes its value and sign across the gapless point. In stark contrast to the $s^*$-wave and $d+{\rm i}d$ cases presented in the main text [Fig.~\ref{fig3}], the energy gap here is symmetric to the Dirac points, and the Chern number is inversion symmetric. To demonstrate the topological properties of the chiral $p_x+ {\rm i}p_y$ state, the quasiparticle spectrum on open lattice at $\mu=-0.95$ is plotted in Fig.~\ref{figA5}(b). Since it is a topological superconductor with $C=-2$, two pairs of gapless states traverse the gap.

\section{The DQMC sign problem in the kagome-lattice Hubbard model}

\begin{figure}[htbp]
\centering \includegraphics[width=8.5cm]{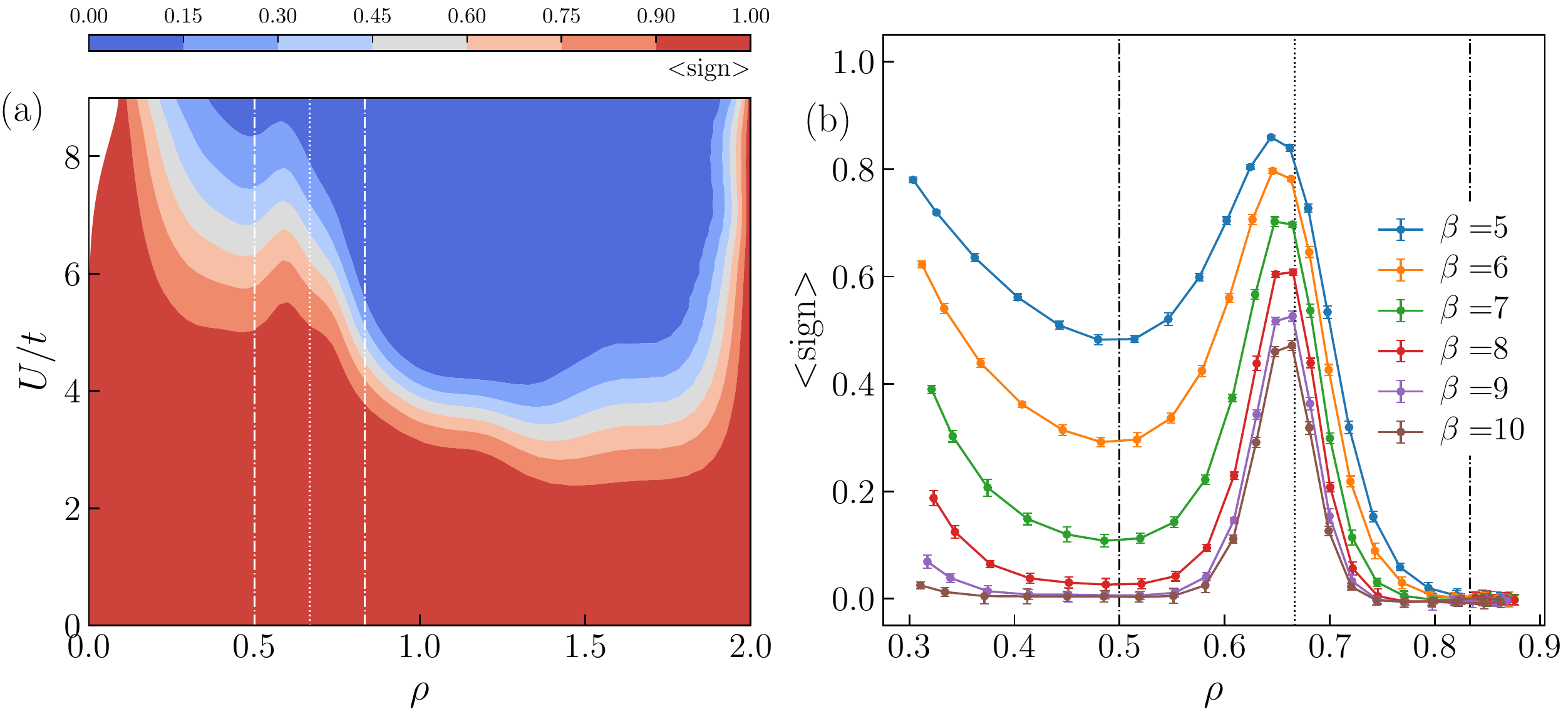} \caption{(a), The contour plot of the average sign in the $(\rho,U)$ plane at $\beta=6$ on a lattice with $N=54$ sites. (b), The average sign as a function of $\rho$ for $U/t=6$ at various inverse temperatures on a $L=9$ kagome lattice. The dotted (dashed-dotted) vertical line marks the position of the Dirac point (the VHSs).}
\label{figa6}
\end{figure}

The kagome lattice is nonbipartite, and has no particle-hole symmetry, thus the DQMC calculations are limited by the sign problem at all densities. Generally, the mean value of the sign is $\langle s\rangle\propto e^{-\beta L\delta f}$~\cite{PhysRevB.41.9301,PhysRevLett.94.170201}, which decreases exponentially with increasing the lattice size $L$ and the inverse temperature $\beta$. We first fix the lattice size, and investigate the evolution of the sign problem with the interaction strength and the inverse temperature. Figure \ref{figa6}(a) presents the landscape of the average sign for the interaction magnitudes ranging from $U/t=0$ to $9$, at all densities in our single-band model. As expected, the sign problem only becomes severe at large interactions and low temperatures. Besides, the average sign is $\langle s\rangle=1$ for both empty and fully-filled bands. At a fixed interaction, $\langle s\rangle$ goes down slowly away from $\rho=0$, but it falls steeply near $\rho=2$ when the Fermi level is in the flat band. As shown in Fig.\ref{figa6}(b), the average sign becomes almost zero in a finite range near each VHS filling at low enough temperatures. Interestingly, $\langle s\rangle$ takes a maximum value between the two VHS points, and the optimal filling approximately corresponds to the location of the Dirac points. Thus DQMC can be performed to larger interactions and lower temperatures near the Dirac points, providing a unique opportunity for DQMC to explore the physics of the correlated Dirac fermions.

\section{More DQMC results}

\begin{figure}[htbp]
\centering \includegraphics[width=8.cm]{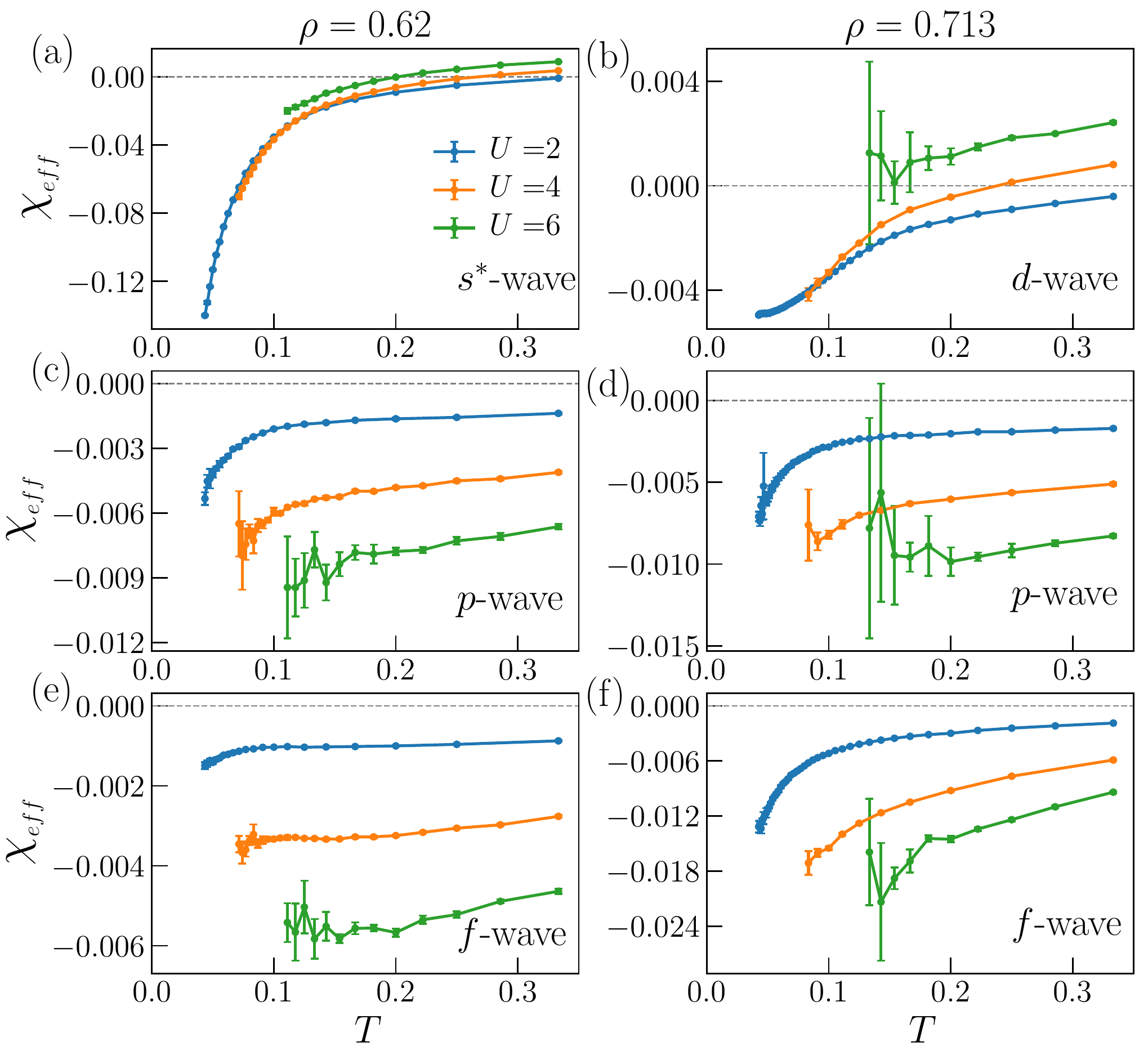} \caption{The effective pairing susceptibility of other pairing channels: (a) $s^*$-wave, (c) $p$-wave, (e) $f$-wave at $\rho_{_1}=0.62$; (b) $d$-wave, (d) $p$-wave, (f) $f$-wave at $\rho_{_2}=0.713$.}
\label{figa8}
\end{figure}

In the main text we have ostensibly focused on the $s^*$($d$)-wave pairing symmetry in electron(hole) doping the Dirac point fillings. Here we provide further justification, by investigating other symmetry channels, starting with Fig.~\ref{figa8}, in complement to Fig.~\ref{fig4} in the main text.

Figure \ref{figa8} shows the effective susceptibility for $p$- and $f$-wave triplet pairings; $\chi_{\rm eff}^{p,f}$ is increasingly negative as the temperature is lowered. Besides, the values of $\chi_{\rm eff}^{p,f}$ decreases as $U$ is increased, suggesting these symmetries are suppressed. The singlet $s^*$($d$)-wave pairing in the hole(electron)-doped case exhibits similar trend with the temperature. Although $\chi_{\rm eff}^{s^*(d)}$ increases with $U$, the values are much smaller than those of the dominating pairing at low temperatures, implying these latter pairing channels are not favored by the interaction.

\begin{figure}[htbp]
\centering \includegraphics[width=8.cm]{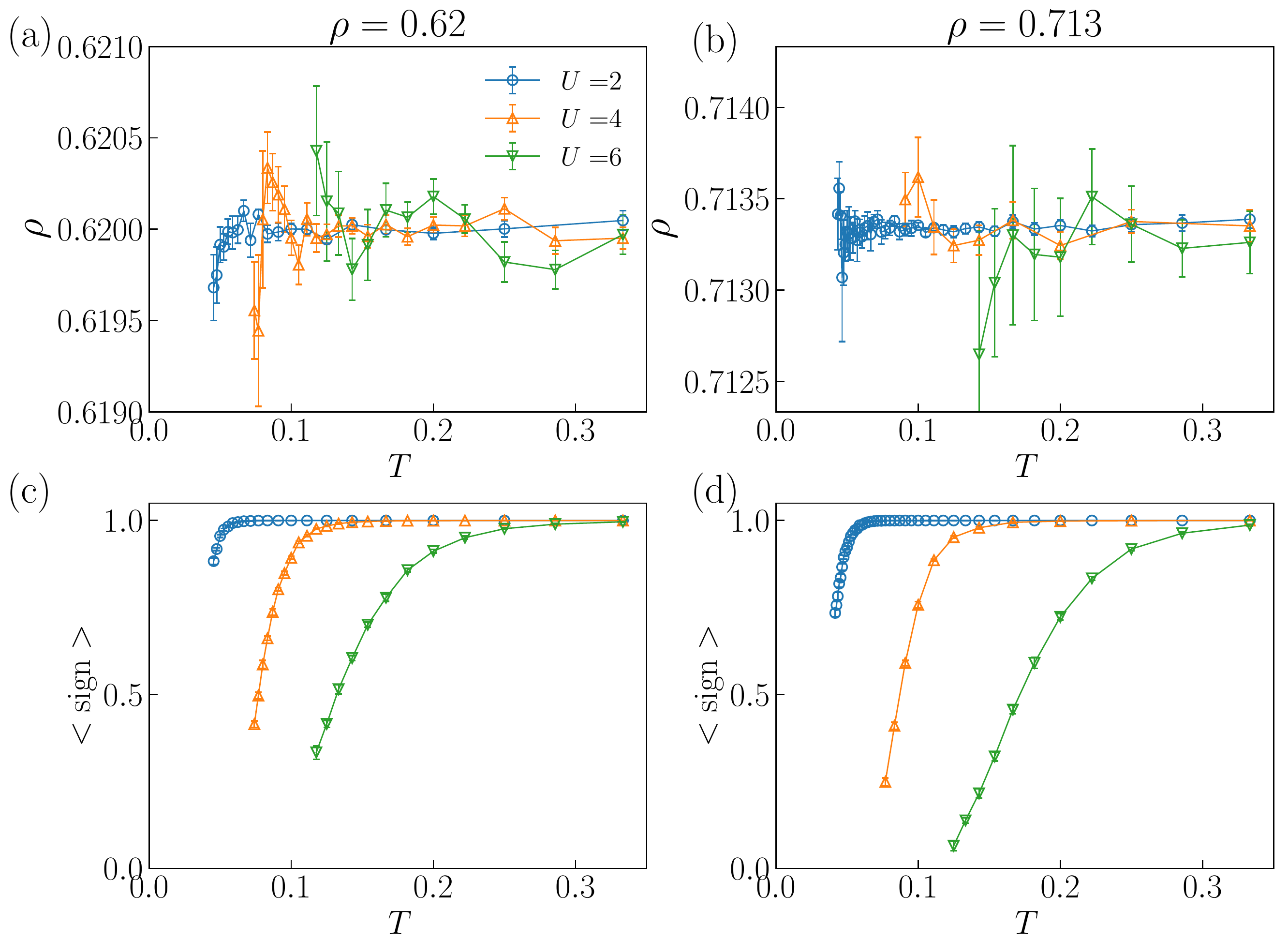} \caption{The average density at the manually determined chemical potential targetting a fixed density: (a) $\rho=0.62$ and (b) $\rho=0.713$. (c) and (d) are the corresponding average signs of (a) and (b), respectively. }
\label{figa9}
\end{figure}

The finite-temperature DQMC method works in the grant canonical ensemble, with the average density being controlled by the chemical potential. Usually $\mu_{\rho}$ (the average density is $\rho$ at $\mu$) is a function of the temperature and the interaction, i.e., $\mu_{\rho}(T,U)$. Thus the chemical potential corresponding to a fixed density has to be found for each set of parameters $(T,U)$. In practice, we perform DQMC calculations with equally-spaced chemical potentials near the desired density. Then from a $\rho-\mu$ curve fitting, the wanted $\mu$ is determined from interpolation. Figures \ref{figa9} (a) and \ref{figa9} (b) plot the average density corresponding to the chemical potentials found using the above procedure for various $U$ at different $T$. The chemical potentials for small interactions and relatively high temperatures can be determined very accurately. However $\rho$ at large $U$ and low $T$ has a relatively large error bar, owing to the severe sign problem in this regime. As shown in Figures \ref{figa9} (c) and \ref{figa9} (d), the average sign begins to drop quickly from a critical temperature. For larger $U$, a clear drop of the average sign happens at higher temperature. These results are consistent with the general rule of the sign problem, i.e., becoming worse at large $U$ and low $T$~\cite{PhysRevB.41.9301,PhysRevB.92.045110}. In our calculations, we have probed the lowest possible temperature for each $U$, average sign permitting.

\begin{figure}[htbp]
\vspace{0.54cm}
\centering \includegraphics[width=8.cm]{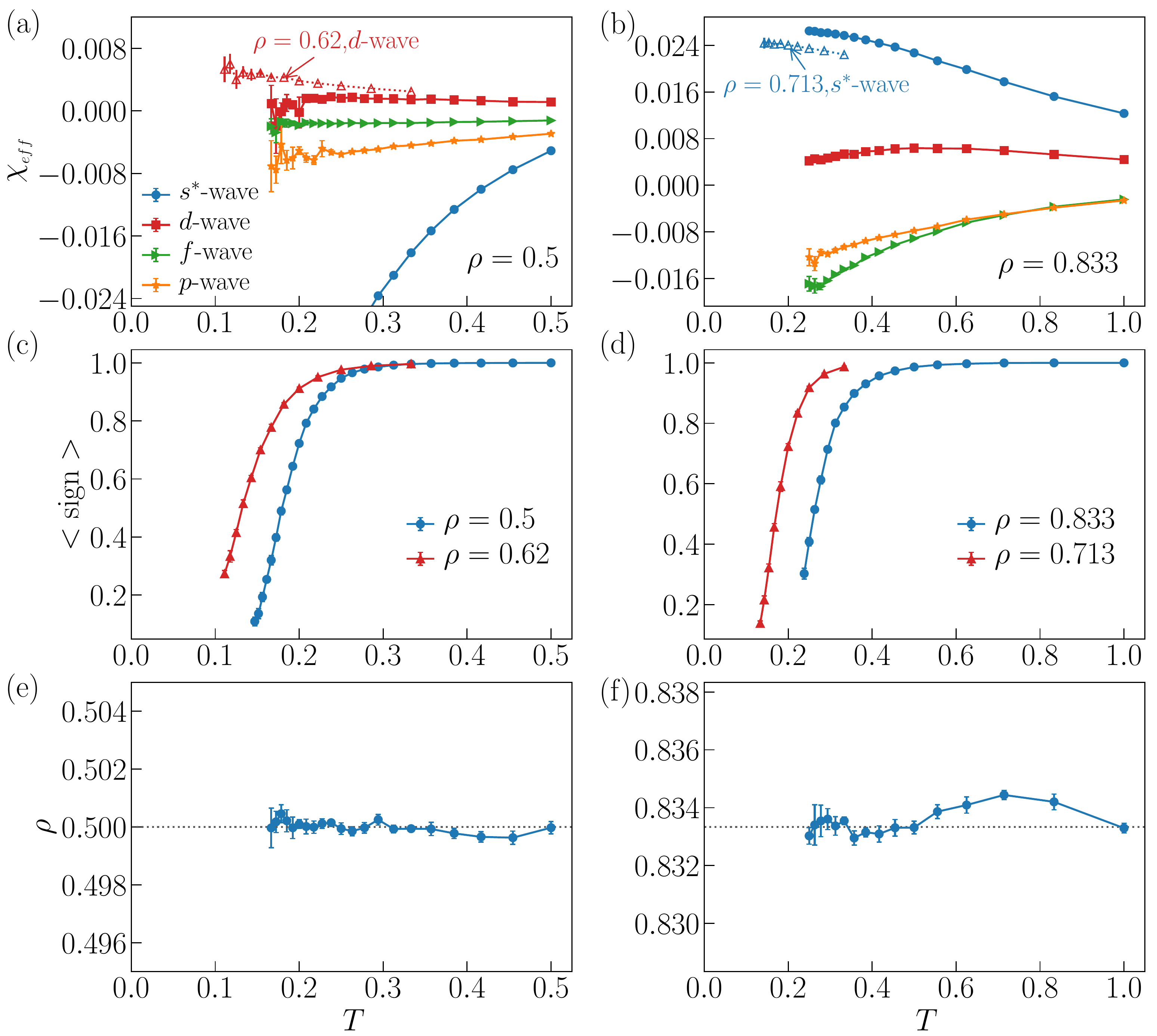} \caption{The effective pairing susceptibility of all allowed pairing
channels: (a) $\rho=0.5$; (b) $\rho=0.833$. (c) and (d) are the corresponding average
signs of (a) and (b), respectively. The average sign at $\rho=0.62$ ($\rho=0.713$) is also plotted in (c) [(d)] for comparison. The average density at the manually determined chemical potential targeting a fixed density: (e) $\rho=0.5$ and (f) $\rho=0.833$. Here the Hubbard interaction is $U/t=6$.}
\label{figa10}
\end{figure}

Figure \ref{figa10}(a) and \ref{figa10} (b) plot the effective susceptibilities for all allowed pairing channels at the VHSs $\rho=\frac{1}{2}$ and $\rho=\frac{5}{6}$, respectively. The values of the $d$($s^*$)-wave channel at $\rho=\frac{1}{2}$ ($\frac{5}{6}$) still dominate. While $\chi^{d}_{\rm eff}$ at $\rho=\frac{1}{2}$ drops quickly and becomes negative at low temperature, $\chi^{s^*}_{\rm eff}$ at $\rho=\frac{5}{6}$ has a  substantial enhancement. It implies that the $d$-wave pairing is destructed at the lower VHS, and the $s^*$-wave channel persists to the upper VHS. The corresponding average signs are shown in Fig.~\ref{figa10} (c) and \ref{figa10} (d), which begin to drop quickly from a higher temperature than that in a more lightly doped case. For the on-site Hubbard interaction $U/t=6$ used in the figure, the temperature accessed is relatively high, which is $T>t/5 ~(t/4)$ at $\rho=\frac{1}{2}~(\frac{5}{6})$. We also plot the average densities corresponding to the manually determined chemical potentials targeting the VHS densities at various temperatures. As shown in Figs.\ref{figa10} (e) and \ref{figa10}(f), the chemical potentials are well controlled to obtain the desired average densities.

\begin{figure}[htbp]
\vspace{0.54cm}
\centering \includegraphics[width=8.cm]{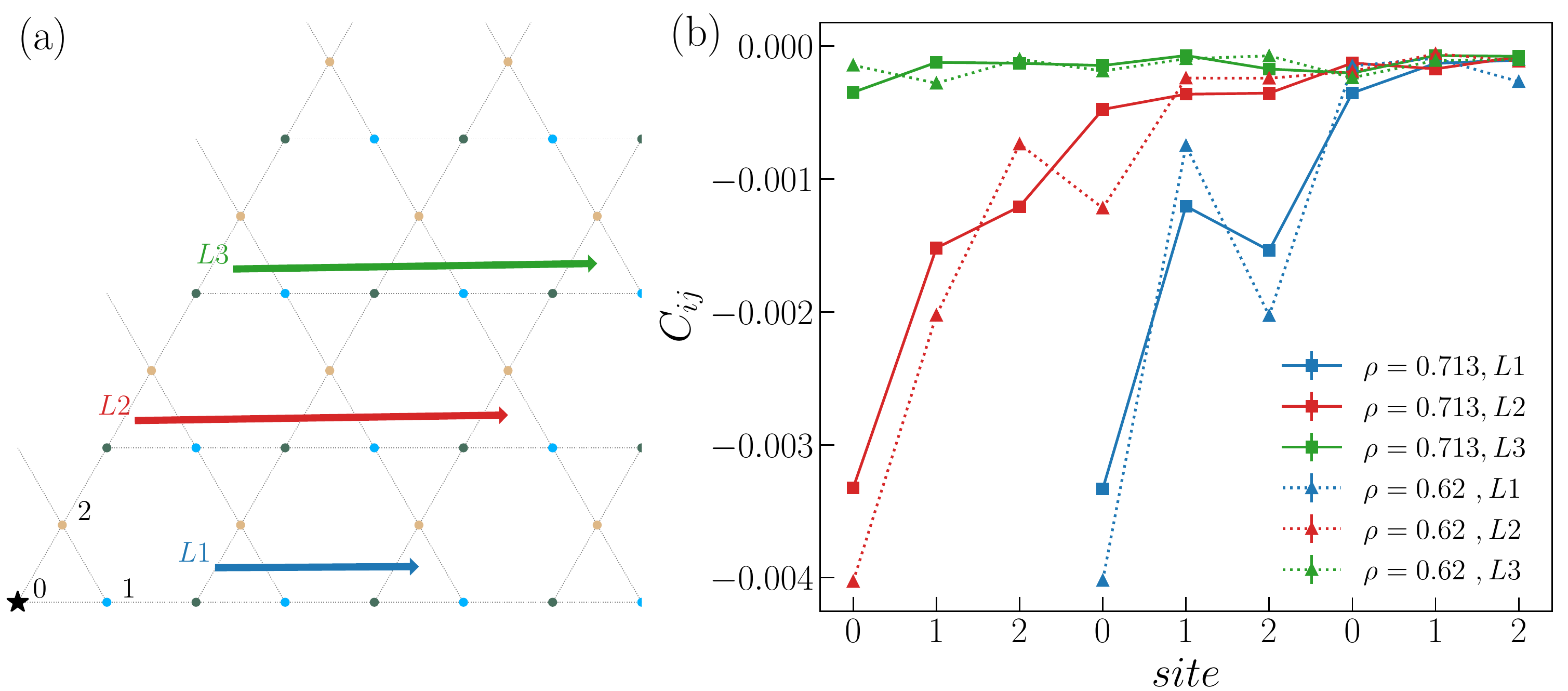} \caption{(a) The $L=6$ kagome lattice on which the DQMC simulations are performed. The star symbol represents the reference site. The colored lines with arrows mark the sites having nonequivalent distances with the reference site. (b) The charge correlations along the paths shown in (a). Here the Hubbard interaction is $U/t=4$, and the inverse temperature is $\beta t=6$.}
\label{figa11}
\end{figure}

Figure \ref{figa11} plots the density-density correlations at $\rho=0.62$ and $0.713$, which is defined as
\begin{align}
   C_{ij}=\langle n_in_j \rangle- \langle n_i \rangle\langle n_j \rangle,
\end{align}
with $n_i=n_{i\uparrow}+n_{i\downarrow}$ the total density on site $i$. In Fig.\ref{figa11} (b), only the charge correlations for nonequivalent pairs of sites are shown. At both fillings, the values of $C_{ij}$ are pretty small, and decrease quickly to zero as $i$ and $j$ depart away from each other. Besides, the values of $C_{ij}$ at the two different fillings differ little from each other. These results suggest the charge fluctuations do not cause the asymmetry of the superconductivity, thus may not account for the superconducting pairing.

\bibliography{ddirac}

\end{document}